# Using various machine learning algorithms for quantitative analysis in LIBS technique

Mohsen Rezaei [a], Fatemeh Rezaei [b*], Parvin Karimi [c]

a) Department of Industrial Engineering, University of Science and Technology of Mazandaran, Behshahr, Iran

b) Department of Physics, K. N. Toosi University of Technology, Tehran, Iran

c) Department of Physics, South Tehran Branch, Islamic Azad University, Tehran, Iran

## Author declarations

## Conflict of interest

The authors declare that they have no known competing financial interests that could have appeared to influence the work reported in this paper.

## Abstract

Laser induced breakdown spectroscopy (LIBS) technique is employed for quantitative analysis of aluminum samples by different classical machine learning approaches. A Q-switch Nd:YAG laser at fundamental harmonic of 1064 nm is utilized for creation of LIBS plasma for prediction of constituent concentrations of the aluminum standard alloys. In current research, concentration prediction is performed by linear approaches of support vector regression (SVR), multiple linear regression (MLR), principal component analysis (PCA) integrated with MLR (called PCA-MLR) and SVR (called PCA–SVR), and as well as nonlinear algorithms of artificial neural network (ANN), kernelized support vector regression (KSVR), and the integration of traditional principal component analysis with KSVR (called PCA–KSVR), and ANN (called PCA-ANN). Furthermore, dimension reduction is applied on various methodologies by PCA algorithm for improving the quantitative analysis. The results presented that the combination of PCA with KSVR algorithm model had the best efficiency in predictions of the most of elements among other classical machine learning algorithms.

*Corresponding author*: fatemehrezaei@kntu.ac.ir

## I. Introduction

Laser induced breakdown spectroscopy is an analytical technique which correlate the spectral signal to the concentration of the analyte according to different mathematical calculations [1]. LIBS is a simple analytical method to identify the elemental composition which uses a focused high energy laser pulse for generation of a plasma from the solid, liquid, or gaseous samples. Multivariate classical machine learning algorithms as the new methodologies have attracted a lot of interest for quantitative analysis in LIBS spectroscopy in recent decades. LIBS technique as a powerful method is an online and fast kind of atomic emission spectroscopy for concentration prediction. Here, the spectroscopic analysis of the light emitted by laser produced plasma is used for identification of the constituent elements of the analyzed sample [2-4].

Different research group investigated on multivariate study in LIBS spectroscopy by different techniques of artificial neural networks (ANN) algorithm [5-9], PCA method [10-12], SVR [13-15], and MLR technique [16-18]. For instance, Unnikrishnan et al [19] have employed Principal Component Analysis (PCA) for the classification of four widely used plastics in LIBS spectroscopy. They have shown that the 375–390 nm region of the LIBS spectra illustrated good results in comparison to other regions without much of the pre-processing. In addition, Ferreira et al [20] have used Artificial neural network for calibration strategy in LIBS technique, aiming to Cu determination in soil samples. They have presented the adequate LOD by utilizing a portable LIBS instrument. Moreover, Dong et al [21] have explored the carbon contents in coal samples by laser-induced breakdown spectroscopy (LIBS) by multiple linear regression (MLR) algorithm, the partial least squares regression (PLSR), and support vector regression (SVR). They have illustrated that the combination of carbon atomic and molecular spectra with both of PLSR and SVR correction improved the quantitative analysis, and the SVR correction helped in reaching better measurement accuracy.

This study represents a combination of LIBS emission spectra with different prediction models. In the current work, the focus is improvement of the precision of the quantitative analysis in LIBS spectroscopy by introducing the best multivariate methodology. Here, a comparison is made between eight multivariate algorithms of MLR, SVR, KSVR, ANN, PCA-MLR, PCA-SVR, PCA-KSVR, and PCA-ANN in terms of both accuracy and precision based on LIBS. To the authors' knowledge, the combination of PCA model with other statistical methods is presented for the first time in LIBS spectroscopy for predictions purposes which caused impressive results. All of these methods are used to quantify the corresponding components of Si, Fe, Cu, Zn, Mn and Mg in seven aluminum standard samples. Mean squared error (MSE), and the mean absolute error (MAE) are employed to evaluate the prediction ability of the mentioned statistical models which are indications of prediction's concentrations.

## II. Experimental set-up

A typical utilized experimental setup of LIBS spectroscopy is presented in Fig. 1 [22-23]. Here, a Q-switched Nd:YAG laser at fundamental harmonic of 1064 nm wavelength, and 10 ns pulse duration, with repetition rate of 10 Hz, and laser energy of 50 mJ is used for plasma creation. The samples are different aluminum standards (1100 series) supplied from the Razi metallurgical research center in Iran. At a particular delay time, the laser light is conducted to a beam splitter and divided into two sections. One part is guided to photodiode for launching the delay generator. Then, the ICCD camera get a pulse from delay generator for beginning the acquisition data. The other part is passed through a λ/2 plate and a Glan–Taylor prism for changing of the laser energy. The focusing of laser pulse is performed by a lens with 20 cm focal length. Moreover, that the position of the strike of the laser pulse is adjusted by a XYZ stage during the whole of experiment. Spatially integrated plasma emissions are collected utilizing a quartz lens with the help of an objective lens and then, are sent to an optical fiber. At next stage, plasma radiations are guided to an Echelle spectrograph (Kestrel, SE200) to receive spectrally resolved light spectrum. During adjustment of the gate and delay time of the ICCD camera (Andor, iStar DH734), the recorded spectral emissions can be temporally studied. For the spectral analysis, the acquisition delay time between the laser pulse and the beginning of the acquisition is changed. Then, the optimum delay time is selected for experimental analysis in order to maximize the signal-to-background ratio (SNR) of the spectral line. In current research, for each sample, 87 LIBS spectra are extracted for each irradiated spot which is repeated 10 times and averaged (in total, 870 spectra per sample) for prediction of constituent concentrations of the test aluminum standard alloys.

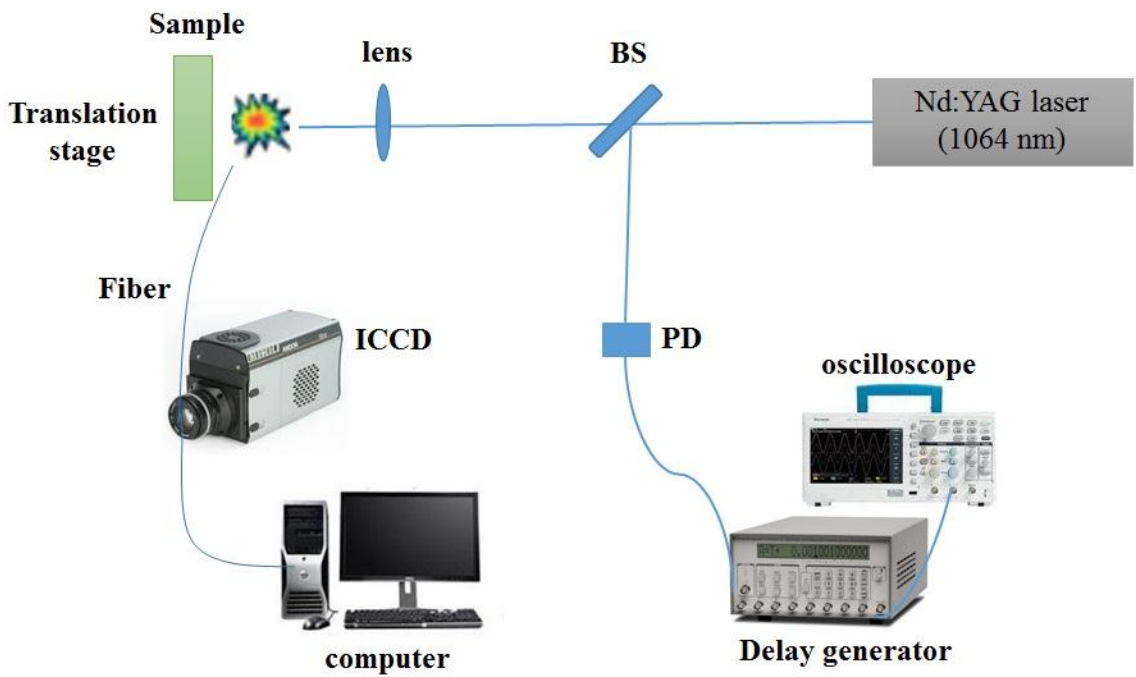

FIG. 1. A Schematic of LIBS experimental set-up.

## III. Statistical prediction methodologies

There are different machine learning algorithms for prediction of the unknown quantities. In this research, the proposed methodologies employed 870 spectra of different standard aluminum samples in correspondence of different concentrations of Si, Fe, Cu, Al, Zn, Mn and Mg. It should be mentioned that due to high concentrations of aluminum elements in all of aluminum standard samples which induce self-absorption phenomenon, they are not utilized for prediction. It can be noted that all the chemometrics codes explained in this section including MLR, ANN, SVR, KSVR, and integrated PCA are calculated by MATLAB 2019b (MathWorks). The details of these algorithms are explained in below sections.

### A. Multiple linear regression (MLR)

Multiple Linear Regression (MLR) method is a statistical multivariate technique which establishes a correlation between the independent and dependent variables (or criterion). MLR is an extensively coupled chemometrics method with LIBS spectroscopy which follows a linear predictor function as below [21]:

$$y = a_0 + a_1 X_1 + a_2 X_2 + \cdots + a_N X_N + \varepsilon \tag{1}$$

where, $a_0$, $a_1$, $a_N$ are the regression coefficients, $\varepsilon$ is the residual error, and $y$ is the dependent variable in a way that this equation must be written for all $M$ samples.

In current research, for designing a MLR model, spectral intensities at different concentrations are considered as independent variables, while elemental concentrations are supposed as dependent parameters. MATLAB software is utilized for calculation of the regression equation and analyzing of the results.

### B. Support vector regression (SVR)

Support vector regression (SVR) methodology is a kernel-based regression method which act according to the principle of support vector machine (SVM). SVR method is introduced as a powerful algorithm for function estimation and pattern recognition. Here, n points with coordinates of $(x_1,y_1)$, ….$(x_n,y_n)$, where, $x_i$ denotes the input spectrum, and $y_i$ indicates the intensities corresponds to the target value, are used as training dataset which n is related to the number of samples. In SVM algorithm, the hyperlane act as a separating line between two data sets, but in SVR method, the line is exploited for prediction of the continuous output. In this case, the margin is a region bounded between two hyperplanes. The main goal is reduction of error, individualizing the hyperplane which maximizes the margin by taking into account that part of the error is tolerated [24].

It can be stressed that SVR algorithm maps the dataset from the nonlinear low dimensional space to linear high-dimensional with application of kernel function so that nonlinear data changes into the linear data in a new coordinate. Hence, the SVR method change the nonlinear relationship of input datasets with using of different kernel functions. The main aim of SVR is finding a function f(x) with a deviation from $y_n$, for each training point $x$, by a value not greater than $\varepsilon$. It should be stressed that each hyperplane can be written as the set of points $x$ implying as flat as possible f(x) function [25]:

$$f(x) = wx + b. \tag{2}$$

here, $b$ is the bias term and $w$ is the normal vector to the hyperplane. Generally, SVR regression algorithm is formulated so that minimize the below functional equation as [26,27]:

$$\frac{1}{2}\|w\|^2 + C\sum_{i=1}^{l}(\xi_n + \xi_n^*). \tag{3}$$

subject to [27]:

$$
\begin{cases}
y_n - (wx_n + b) & \leq & \varepsilon + \xi_i \\
(wx_n + b) - y_n & \leq & \varepsilon + \xi_i^* \\
\xi_n, \xi_n^* & \geq & 0
\end{cases} \tag{4}
$$

where, $\xi_n$ and $\xi_n^*$ are two positive slack variables for measuring the deviation. $C$ is a box constraint constant with a positive value which controls many of imposed on observations placed outside of epsilon margin ($\varepsilon$) and help to prevent overfitting.

It should be mentioned that linear SVR in dual formula follows a Lagrangian function built from primal function by regarding to the nonnegative multipliers $\alpha_n$ and $\alpha_n^*$ for every observation $x_n$ as:

$$y = \sum_{i=1}^{l}(\alpha_n - \alpha_n^*) \cdot (x_n x) + b. \tag{5}$$

### C. Kernelized support vector regression (KSVR)

Generally, SVR can be studied by a linear or nonlinear regression algorithm according to the used kernel function. It can be stressed that the Kernelized support vector regression (KSVR) method is faster than the standard SVR in the non-linear regression related to large datasets, while produce the high correctness in the prediction. SVR can support the nonlinear relationship of input datasets with different kernel functions [15].

In non-linear SVR, the data will be transformed into a higher dimensional feature space by kernel functions ($K(x_n, x)$), for providing the linear separation as bellows [28]:

$$y = \sum_{i=1}^{l}(\alpha_n - \alpha_n^*) \cdot K(x_n, x) + b. \tag{6}$$

Kernel functions can be explained by a polynomial function as:

$$K(x_n, x) = (x_n.x)^d. \qquad\qquad \text{d=2, 3, ...} \qquad\qquad (7)$$

Or by Gaussian radial basis function as [26]:

$$K(x_n, x_j) = exp\left(-\frac{\|x_n - x_j\|^2}{2\sigma^2}\right). \qquad\qquad (8)$$

where, $\sigma$ is the width of the kernel function, and $x_n$, $x_j$ are the $n^{th}$ and $j^{th}$ support vectors. It can be mentioned that in linear form of SVR, kernel function equals to $x_n x$ in Eq. (6).

## D. Artificial neural network (ANN)

The Artificial neural networks (ANN) as an innovative computational approach has been inspired right from its inception by recognizing that the human brain calculates in a completely different way from the conventional digital computer. Generally, the brain is a severely nonlinear information-processing system and parallel computer. It has the ability to manage its structural constituents (neurons) to analyze certain computations such as perception and motor control, pattern recognition very faster than the fastest digital computer. In fact, features of ANNs as a powerful mathematical technique can be considered by high parallelism, special data processing, acquiring knowledge through learning process, non-linear data mapping, highly weighted connections between elements, adaptability and generalization ability. The most well-known applications of ANNs are indicated as classification, pattern recognition, and predictions in different areas such as psychology or engineering [29]. These drastic simulation tools are comprised of giant number of single units as artificial neurons, topology and learning algorithm.

The behavior of nodes as artificial components is simulated according to natural neurons which can be connected to each other by getting and sending special signal through electrical and chemical alternation. Actually, the neurons in ANN structures are simulated by transfer functions and are organized in input, output and hidden layers. They are jointed together with the random coefficients (weights) which are continuously customized for performing optimization.

In processing of information in a single node, first of all, the arrival weighted activations accompanied by previous nodes are combined together. Then, they pass through a transfer function as a feature of a specific node.

After that, the outcome results created by all the nodes in one layer will be transmitted to the nodes of next layer. This procedure continues to all of layers for attaining the output data.

Figure 2 shows a schematic of a multilayer feed forward network with some features of artificial neurons. It is constructed from main three sections: i) input layer with definite neurons equal to initial signals. ii)  hidden layers placed between first and last layers for enhancement of responses which specific arbitrary nodes. iii) output layer consisted of the neurons which convey equivalent network results.  It can be mentioned that none of the single-layer nodes are jointed together and information can be transferred between layers according to the given weights. The estimation between desired and received results determines the efficiency of ANNs. Actually, to obtain the most optimal responses with the least error, the learning algorithms can be employed.

Learning rules or training processes are explained as a group of input-output matching patterns which endorses to obtain the best results related to the input data. Consequently, the ANNs try to modify random weights as conjunction coefficients to increase the performance of networks through decreasing of error.

The proceeding is dependent upon forward and backward propagation for modification the value of all weights. It can be stressed that one of the most well-known learning instructions is named back-propagation (BP) based on gradient-descent algorithm.

In forward propagation, the whole of connectivity components are constant and final consequences are calculated. In backward dispersion, all values of weights are refined based on the computed error in certain manner which are then used again for the subsequence forward processing. This approach will be utilized continuously to attain the favorable purpose.

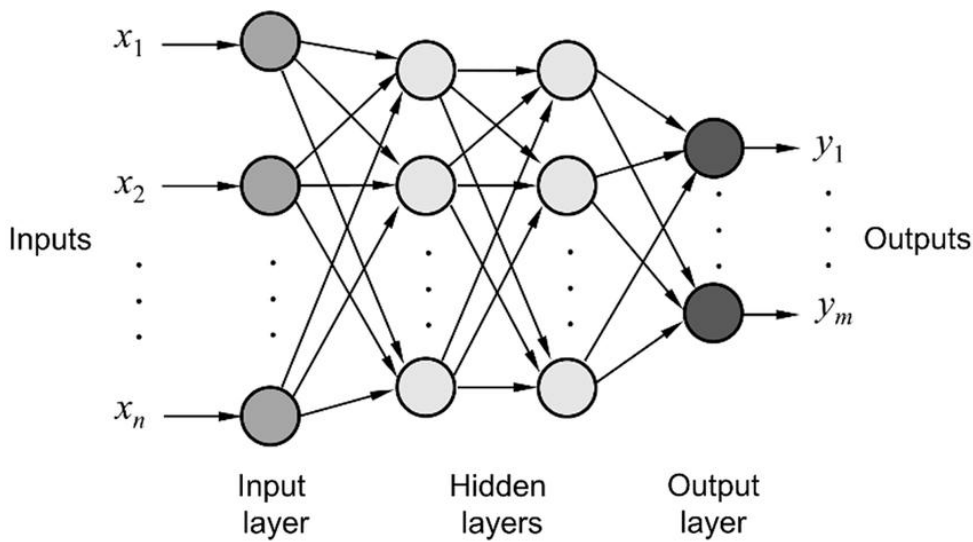

FIG. 2. A diagram of the structure of neural network [30].

## E. Principal component analysis (PCA)

Principal component analysis is a tool for projection of high dimensional dataset into a small number of variables (named principal component, PC), and utilizing them for performance of a change of data basis. PCA is a powerful chemometric methodology which exploit to reduce data dimension by applying an orthogonal transformation. Here, two new matrices are generated: a loading matrix that shows the weight of the original variables, and a score matrix containing the projections of the samples so that samples x variables or raw data matrix are changed into a score matrix. It can be mentioned that a conversion is performed for transformation of correlated variables into a group of uncorrelated variables (principal components, PCs). On the other hand, the principal components (scores plot) are plotted and clusters will be observed in the graph, which are representative of the samples with similar composition/spectrum. The details of this method is given elsewhere in literatures of 25. In this study, PCA is done by using an in house developed Matlab routine (The Mathworks INC., Natick, USA). The raw data matrix is constructed from spectral emissions of different aluminum samples [12].

## F. Integration of PCA with other multivariate methods (PCA-ANN, PCA–SVR, PCA–KSVR, PCA–MLR)

Generally, the integration of traditional principal component analysis with other multivariate methods can increase the prediction accuracy. The strategic analysis is that PCA is used as a pre-processing step during other multivariate modeling for reduction of the dimensionality of the original multivariable dataset. One of the main advantages of coupling PCA with other methodology is reduction of training time in encountering large datasets processing, since, PCA make data compression. Furthermore, for instance, ANN is very sensitive to correlations amongst inputs, therefore in input data with strong correlations, PCA integration can remediate this problem. Consequently, for tackling these mentioned problem and observing the enhancement of the accuracy in a lot of prediction literatures [31], this paper, integrated PCA with ANN (called PCA–ANN), SVR (called PCA–SVR), KSVR (called PCA–KSVR), and MLR (called PCA–MLR) for finding the best models.

## IV. Errors estimation and accuracy evaluation

In statistical analysis, the concept of error  is a fundamental concept for measuring of the effectiveness of an estimator or predictor. It can be noted different mathematical relation can be used for assessment of errors. For instance, mean squared deviation (MSD) or mean squared error (MSE) estimates the quality of a predictor as bellows [32]:

$$MSE = \frac{1}{n}\sum_{i=1}^{n}(y_i - \widehat{y_i})^2. \tag{9}$$

In addition, root mean square error (RMSE) evaluates the goodness of the prediction for each trial as [32,33]:

$$RMSE = \sqrt{\sum_{i=1}^{n}\frac{(\widehat{y_i}-y_i)^2}{n}} \tag{10}$$

Mean absolute error (MAE) is also computed by considering the average of all absolute errors of the results as [34]:

$$MAE = \frac{1}{n}\sum_{i=1}^{n}|y_i - \widehat{y_i}|. \tag{11}$$

here, $y_i$ and $\widehat{y_i}$ are the target and estimated concentration magnitudes corresponds to spectrum $i$, and $n$ is the number of test spectra taken into account.

## V. Results and discussion

The most frequent methods for composition prediction in LIBS spectroscopy are calibration curve and artificial neural network approaches. In this work, quantitative determinations are carried out by using multivariate method of linear approaches such as MLR [35], PCA-MLR [36], SVR [15], and PCA-SVR [37], and nonlinear methods such as ANN [38], KSVR [39], PCA-ANN [40], and PCA-KSVR [41].

A feed-forward perceptron ANN with back-propagation algorithm is used for data prediction. In all of calculations, MATLAB software is utilized and ANN toolbox is employed for model developing. Here, one sample is selected into the validation set for every seven samples, and the rest samples are used as training set. It can be stressed that the data from 5 physical samples is used for training and 1 for validation. This have been repeated by randomly selecting the 1 validation sample from the 6 samples, carrying out 6 iterations in total. After optimizing each regression model in this way, the last (seventh) sample have been used for testing. Levenberg–Marquardt is the most widely learning algorithm which is used in this research for network training [42]. Indeed, it can be stressed that based on different literatures [29,43], increasing the number of hidden layers induces the "overtarining" problem, therefore, it is seldom necessary to consider more than one hidden layer. The performance of the ANN model is shown in table 1 for a typical Fe element versus different numbers of hidden neurons. As shown in this table, the best ANN model is obtained with just one neuron in the hidden layer which is highlighted with red color. It can be noted that this process can be repeated in a similar way for other elements.

Table 1. Determination of the number of the used hidden neurons for Fe element.

| Number of hidden neurons | MSE | MAE |
|:---:|:---:|:---:|
| **1** | **0.016** | **0.093** |
| 2 | 0.021 | 0.110 |
| 3 | 0.029 | 0.113 |
| 4 | 0.033 | 0.149 |
| 5 | 0.040 | 0.159 |
| 6 | 0.040 | 0.163 |
| 7 | 0.033 | 0.148 |
| 8 | 0.066 | 0.215 |
| 9 | 0.090 | 0.242 |
| 10 | 0.077 | 0.213 |

In artificial neural network algorithm, the trial and error method is employed for choosing the best transfer functions. Here, different transfer functions of purelin (linear), logsig (Log-sigmoid), and tansig (tangent sigmoid) [44,45] are tested for all of the elements in order to attain the best prediction with minimum errors. It is well known that the observed errors in the neural network analysis is dependent on the initial random magnitudes of the neural weights. It can be stressed that in this research, the calculations are repeated 30 times for each network and the error of the network is actually reported as the average of MSEs, and MAEs over whole executions. A summary of the best transfer function for both of hidden and output layers with optimum number of hidden neurons is presented in table 2. Furthermore, in this table, the related error bars of MSE, and MAE are illustrated for showing the precision of measurements. As it is seen, the results demonstrate that the best performance of the proposed ANN method is happened with different transfer function related to the hidden and output layers for each element. Additionally, it is observed that Fe has the least error values among other elements during using of purelin transfer functions in both of the layers. This fact can be attributed to low concentration of Fe element which obeys a linear trend in the calibration curve.

Table 2. The best transfer functions of the hidden and output layers and the optimum number of hidden neurons based on ANN calculations. Error values of MSE, and MAE represent the precision of utilized methods.

| Elements | Number of hidden neurons | Transfer Functions | | Errors | |
|---|---|---|---|---|---|
| | | Hidden layer | Output layer | MSE | MAE |
| Fe | 1 | purelin | purelin | 0.016 | 0.093 |
| Zn | 10 | logsig | tansig | 0.084 | 0.288 |
| Si | 1 | purelin | purelin | 0.142 | 0.306 |
| Mn | 1 | purelin | purelin | 0.063 | 0.207 |
| Cu | 3 | tansig | purelin | 0.033 | 0.148 |
| Mg | 6 | logsig | logsig | 1.378 | 1.174 |

Fig. 3 presents the results of best predictions for different elements of Fe, Cu, Zn, Si, and Mn with artificial neural network method. As it is clearly seen, the slope of correlation curve between predicted and nominal concentration is near to 1 for all the elements.

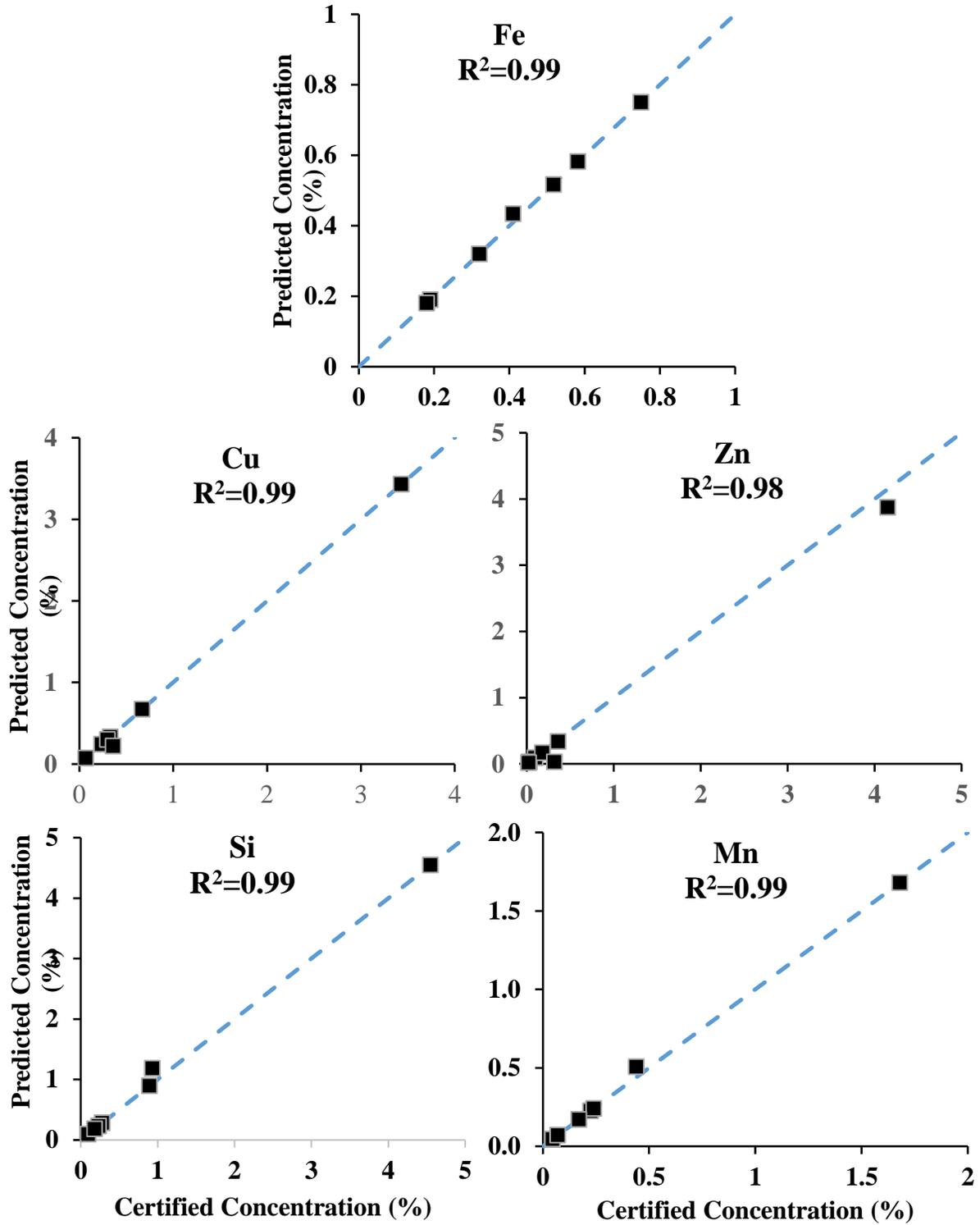

FIG. 3. Concentration prediction for different elements of Fe, Cu, Zn, Si, and Mn by artificial neural network.

The single SVR and MLR models simply utilizes these methods to input variables for forecasting the element concentration without using any PCA, or other preprocessing tools such as linear independent component analysis (ICA) or nonlinear ICA. A comparison of the analytical performances of MLR, and SVR methods is shown in figure 4 for Si element due to its best prediction. Si concentration shows the best prediction by both SVR and MLR models with a low value for MSE and MAE. Here, the prediction models of MLR and SVR approaches as two linear methods are introduced as quantitative models which use the composition and spectral intensities of aluminum's alloys for developing their performance. The correlation between the certified and predicted concentrations by these two models in this research represents a good quantitative measure of their prediction. It is obvious that the analytical predictions of the two models are very similar for Si element. Moreover, after calculations for other elements, it is clearly seen that in both approaches, sometimes great deviations are seen in high concentrations which is due to self-absorption effects which causes the multivariate methods could not perform well in prediction of aluminum's composition [46]. Besides, comparison of two methods of SVR and ANN shows that the SVR model is more accurate than ANN in predicting aluminum contents except for Zn and Fe elements. Additionally, ANN forecasts more exact concentrations compared to MLR approach except for Si element.

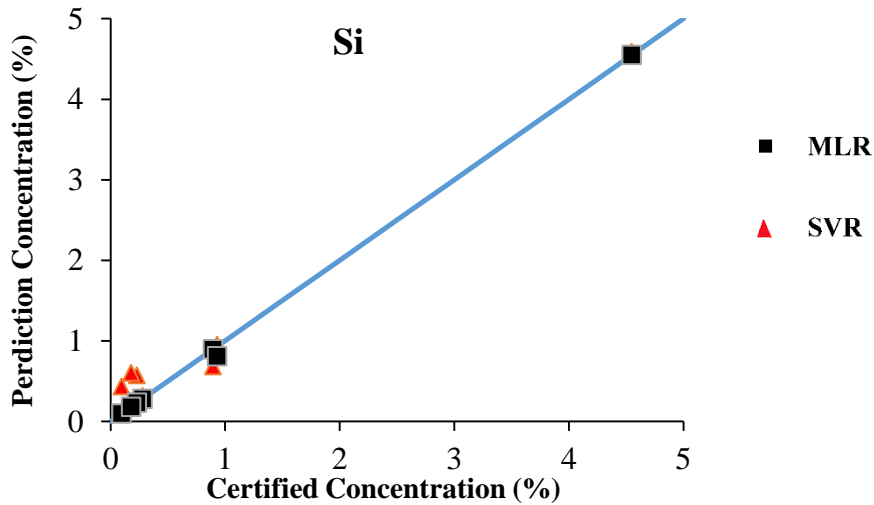

FIG. 4. A comparison between concentrations prediction by MLR and SVR multivariate methods for Si element with coefficient of determination of $R^2$=0.99 and $R^2$=0.96, respectively.

Another approach for prediction of the composition of the constructed aluminum alloys is using kernel-based machine learning methods (KSVR). The influence of different kernel functions on composition prediction is illustrated in table 5 and their performance is compared in terms of MSE, and MAE. As it is seen in table 3, the Gaussian kernel almost forecasts better results rather than other kernels functions in different elements which is in good agreement prediction with results reported in Refs [15,47]. It is worth noting that linear kernel produces the worst prediction performance for most of elements. Additionally, it could be mentioned from the KSVR results that the error values decreased a lot in all elements which induce that integration with kernel function improved a lot the SVR analysis performance. This fact proposes that KSVR which comprises the kernels function provides better performance than single SVR method. Red colors in this table show the lowest values of errors for different elements. Consequently, it can be concluded that the KSVR method is able to reproduce the concentrations with acceptance standard errors.

Table 3. A comparison between different kernel functions performance for different aluminum's elements.

| Elements | Linear | | Polynomial | | Gaussian | |
|:---:|:---:|:---:|:---:|:---:|:---:|:---:|
| | **MSE** | **MAE** | **MSE** | **MAE** | **MSE** | **MAE** |
| **Fe** | 0.078 | 0.280 | 0.0002 | 0.013 | **0.0001** | **0.012** |
| **Si** | **0.012** | **0.109** | **0.012** | **0.109** | 0.208 | 0.456 |
| **Mn** | 0.217 | 0.465 | **0.001** | **0.035** | 0.069 | 0.263 |
| **Cu** | 0.031 | 0.175 | 0.230 | 0.480 | **0.002** | **0.045** |
| **Mg** | **0.053** | **0.230** | **0.053** | **0.230** | 0.059 | 0.243 |
| **Zn** | 0.523 | 0.723 | 0.236 | 0.486 | **0.014** | **0.120** |

As an example of best prediction with KSVR approach is depicted in figure 5 for Fe element with correlation of 0.97.

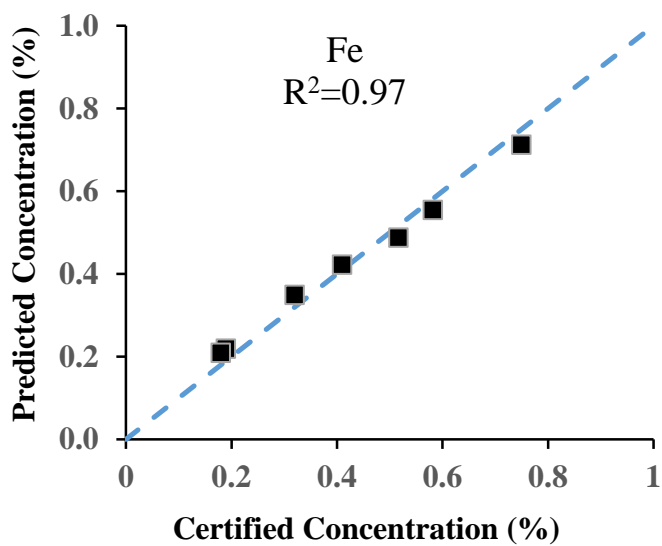

FIG. 5. Prediction of concentration of Fe element with Kernelized support vector regression including correlation of 0.97.

In the next stage, PCA algorithms are added to all of the mentioned methodologies of ANN, MLR, SVR and KSVR approaches. A feature which causes the application of the PCA integration methods in LIBS analysis much simpler (and more precise) with respect to other traditional algorithms is reduction of dimension. In all assimilation with PCA methods, PCA is first applied to the input variables to generate the principal components (PCs) and then, the considered analyses are conducted according to the generated PCs. Again, it can be mentioned that for representation of the characteristics of input data, all PCs would be adopted to be used as new input variables for the mentioned model. More precisely, the proposed integrated approaches can be compared with their related alone methods.

A comparison between performance of two methods of PCA-SVR and PCA-KSVR is presented in figure 6 for two elements of Fe and Mn. Approximately, similar trends are observed by both of approaches.

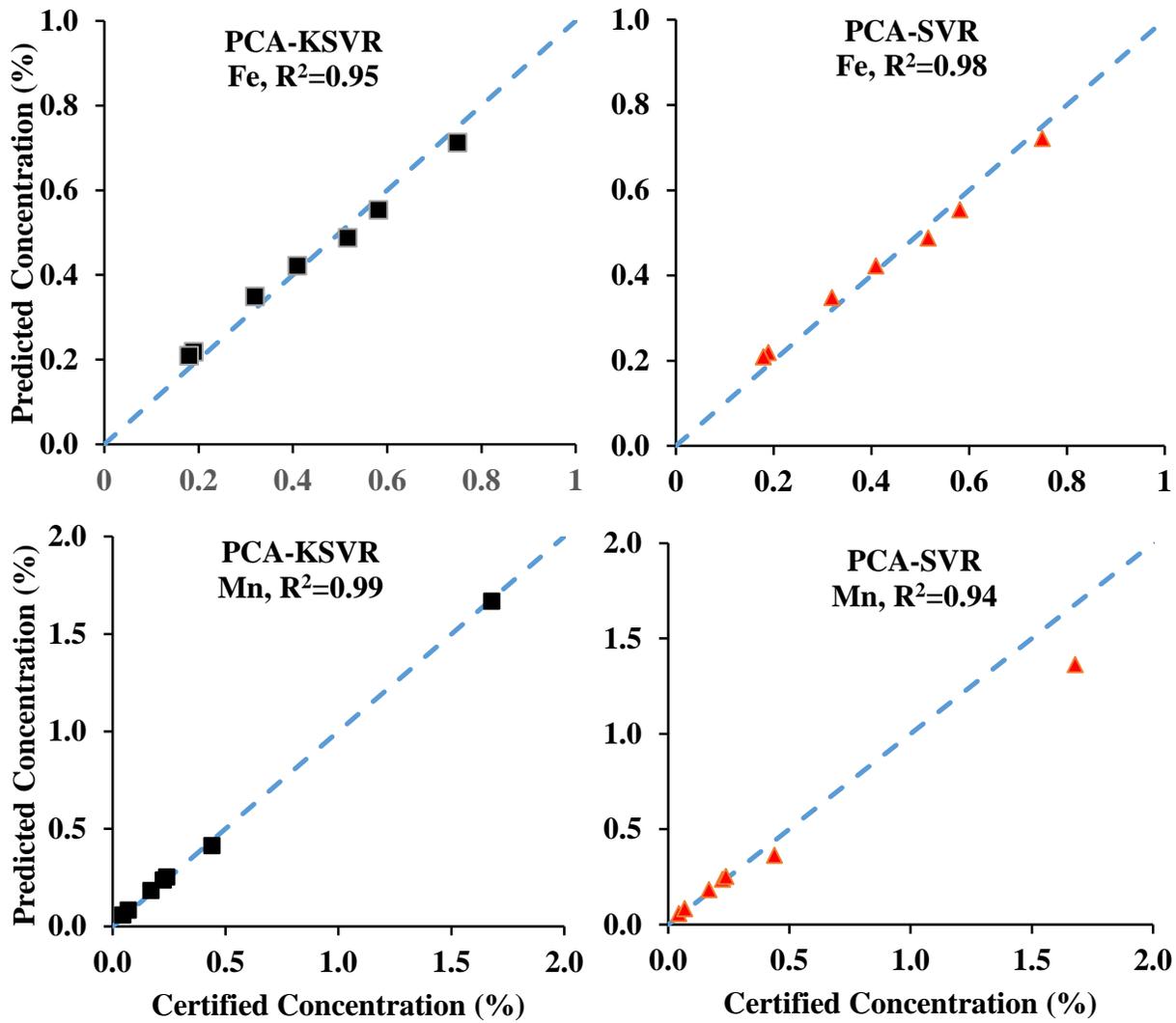

FIG. 6. Concentration predictions for two elements of Fe and Mn with combinational methods of PCA-SVR and PCA-KSVR.

Then, for evaluation the capability of two combinational methods of PCA-ANN and PCA-MLR, in quantitative analysis, Fig.7 is represented for Si, Mg and Cu elements. As it is seen in this figure, the accuracy of PCA-MLR is similar to PCA-ANN in prediction of concentrations while it is not true when PCA technique is not combined with ANN and MLR methods. In this figure, the correlation slope of curve between predicted and certified concentration is close to 1 for all the elements.

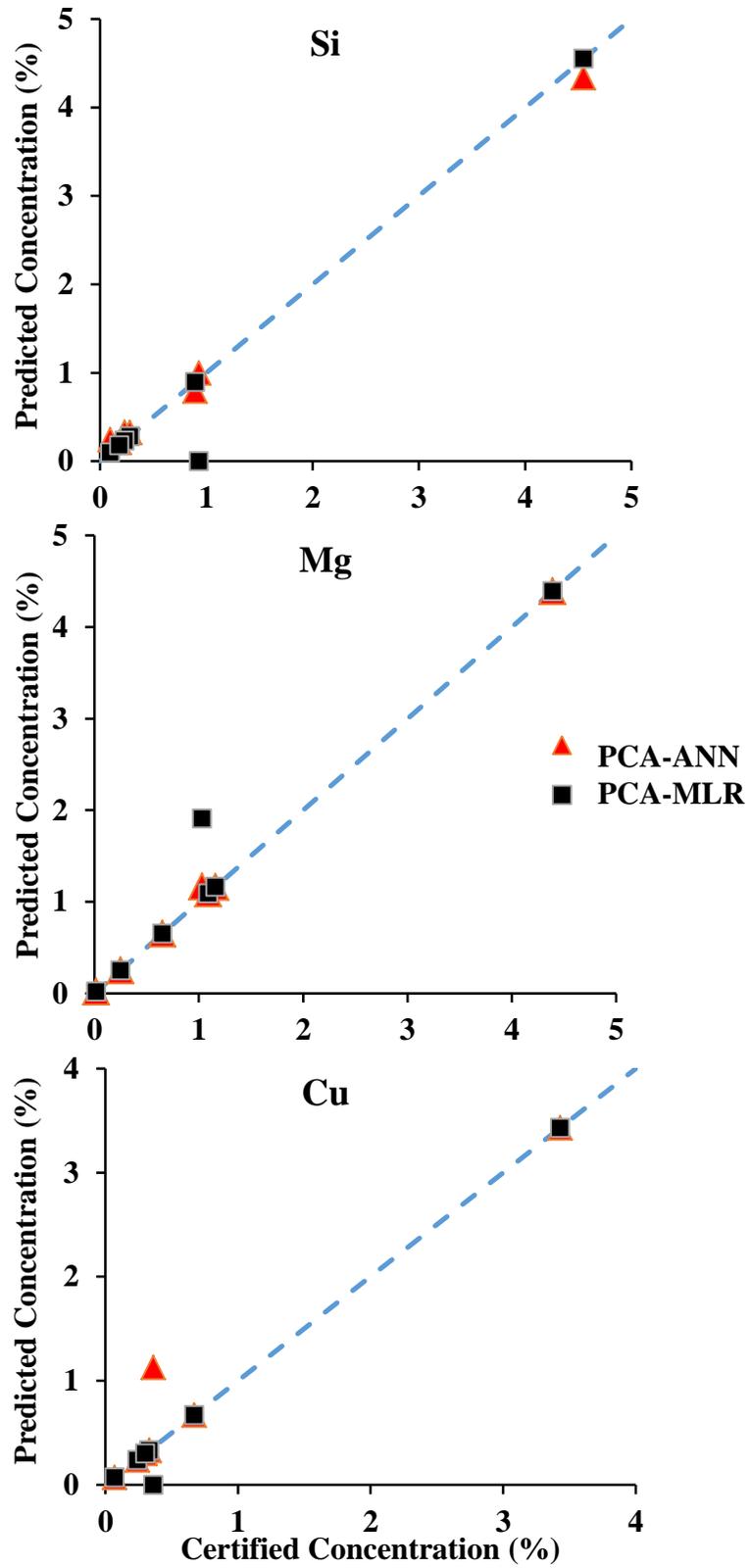

FIG. 7. Concentration prediction of Si, Mg, and Cu elements by two methods of PCA-ANN and PCA-MLR.

As a final consideration, a comparison summarized in table 6 represents the prediction results of mean concentration by eight proposed approaches of ANN, MLR, SVR, KSVR, PCA-ANN, PCA-MLR, PCA–SVR, and PCA-KSVR. The validity of these methods is estimated by comparing the statistical error values of mean MSE, and mean MAE calculated from constituent elements of standard aluminum samples. It can be stressed that the least error values are highlighted with red color. As clearly seen in this table, in most of elements but not all, integration with independent component analysis (PCA) enhanced the accuracy of the single form of that special method. For instance, MLR methodology showed relatively high errors for Mg element, but applying PCA, improved their prediction very much.

The results in table 4 prove the evidence that the proposed PCA-KSVR as the best prediction methodology has produced the significantly lowest MAE, and RMSE (%) for all of elements except for Si. From this table, it is found that the highest error value in the PCA–KSVR achieved to 0.18 % which reflect the robustness of this method. The comparison prediction results affirm that the PCA-KSVR approach not only improves the prediction accuracy of the single SVR method, but also outperforms the competing methods in forecasting. Apparently, PCA-SVR can be introduced as the most accurate methodology after PCA-KSVR prediction method which produce low measurement errors and high accuracy.

Generally, the integration of PCA and machine learning approaches illustrated good performance in different forecasting fields [27], such as prediction of greenhouse gas emissions 48, evaluation of coronary artery diseases [49], and predicting gasoline homogeneous charge compression ignition combustion trend during transient operation [50].
In this table, it is demonstrated that when the PC integration methods are not considered, KSVR approach can better predict the compositions with higher precisions rather than simple MLR, SVR or optimized artificial neural network model.

Table 4. A comparison between different multivariate approaches in prediction of aluminum contents by calculation of mean squared error (MSE), and the mean absolute error (MAE).

| Methods | Fe | | Zn | | Si | | Mn | | Cu | | Mg | |
|---|---|---|---|---|---|---|---|---|---|---|---|---|
| | MSE | MAE | MSE | MAE | MSE | MAE | MSE | MAE | MSE | MAE | MSE | MAE |
| MLR | 0.020 | 0.140 | 0.326 | 0.571 | 0.015 | 0.121 | 1.518 | 1.232 | 1.935 | 1.391 | 4.066 | 2.016 |
| PCA-MLR | 0.079 | 0.280 | 1.447 | 1.203 | 0.865 | 0.930 | 0.087 | 0.295 | 0.130 | 0.360 | 0.768 | 0.877 |
| SVR | 0.029 | 0.171 | 2.413 | 1.553 | 0.0004 | 0.020 | 0.054 | 0.233 | 0.020 | 0.141 | 0.053 | 0.229 |
| PCA-SVR | 0.0002 | 0.012 | 0.0004 | 0.021 | 0.131 | 0.362 | 0.006 | 0.077 | 0.033 | 0.181 | 0.034 | 0.184 |
| KSVR | 0.0001 | 0.012 | 0.014 | 0.120 | 0.012 | 0.109 | 0.0012 | 0.035 | 0.0020 | 0.045 | 0.053 | 0.230 |
| PCA-KSVR | 0.00001 | 0.004 | 0.001 | 0.036 | 0.021 | 0.144 | 0.0007 | 0.027 | 0.0017 | 0.041 | 0.033 | 0.183 |
| ANN | 0.016 | 0.093 | 0.084 | 0.288 | 0.142 | 0.306 | 0.063 | 0.207 | 0.033 | 0.148 | 1.378 | 1.174 |
| PCA-ANN | 0.003 | 0.056 | 1.641 | 0.975 | 1.785 | 1.020 | 0.179 | 0.423 | 1.528 | 0.883 | 1.189 | 0.729 |

Taking into account the above considerations, the least mean squared error (MSE) and the mean absolute error (MAE) of the prediction values for the validation samples are related to Fe element and the worst one are seen in Mg element. This fact is due to the relatively low concentration of Fe and high content of Mg in aluminum alloys compared to other constituent elements, however, in the case of Mg, in order to improve the efficiency and reliability, PCA-KSVR method can provide a great reduction in error calculations.

Finally, the best proposed method is introduced in table 5 for all of elements with representation of the predicted values by these approaches. As it is seen, PCA-KSVR reported the nearest predicted values to certified concentrations for Fe, Zn, Mn, and Cu elements.

Table 5. Best methods for calculation of concentrations of different elements of Fe, Zn, Si, Mn, Cu, and Mg.

| Element | Certified Concentration | Predicted Concentration | Best Method |
|---|---|---|---|
| Fe | 0.41 | 0.414 | PCA-KSVR |
| Zn | 0.32 | 0.341 | PCA-SVR |
| Si | 0.93 | 0.950 | SVR |
| Mn | 0.44 | 0.413 | PCA-KSVR |
| Cu | 0.36 | 0.401 | PCA-KSVR |
| Mg | 1.03 | 0.847 | PCA-KSVR |

## VI. Conclusion

The determination of the exact composition of the aluminum alloys is somewhat difficult by LIBS spectroscopy, due to the important matrix effects and the nonlinearity relation of the spectral intensities versus concentration. In this paper, different statistical prediction methods such as MLR, ANN, SVR, KSVR, PCA-MLR, PCA-ANN, PCA-SVR, and PCA-KSVR as methodological approaches are coupled with LIBS spectroscopy technique in order to evaluate the effectiveness of the proposed models and introduce the best quantitative methods. For each model, the general variation of the errors including MSE, and MAE is reported. It was seen that in most of cases, the assimilation of PCA improved significantly the performance of chosen approach compared to single formation of that model. The experimental results verified that PCA-KSVR is preferable with respect to the other approaches only through the comparison of different error values. Finally, a definite benefit of the proposed approaches is the possibility of using them effectively for giving information on constituent elements of each arbitrary sample in LIBS analysis.

**Acknowledgement**

The authors are thankful from Professor Seyed Hassan Tavassoli for sharing his laboratories equipment and his kind supports.

**Data availability statement**

The data that support the findings of this study are available from the corresponding author upon reasonable request.